\documentclass[twocolumn,superscriptaddress,floatfix]{revtex4}
\usepackage{graphics,epsfig,amsfonts,amssymb,amsmath,ulem,color,xcolor}
\begin{document}
\title{2-dimensional  semiconductors pave the way towards dopant based quantum computing}
\definecolor{ultramarine}{RGB}{12,32,200}
\author{J.C. Abadillo-Uriel}
\affiliation{Materials Science Factory, Instituto de Ciencia de Materiales de Madrid,
ICMM-CSIC, Cantoblanco, E-28049 Madrid (Spain).}
\author{Belita Koiller}
\affiliation{Instituto de F\'isica, Universidade Federal do Rio de Janeiro, Caixa Postal 68528, Rio de Janeiro, RJ 21941-972, Brazil}
\author{M.J. Calder\'on}
\affiliation{Materials Science Factory, Instituto de Ciencia de Materiales de Madrid, ICMM-CSIC, Cantoblanco, E-28049 Madrid (Spain).}
\date{\today}
\begin{abstract}
Since the 1998 proposal to build a quantum computer using dopants in semiconductors as qubits, much progress has been achieved on semiconductors nano fabrication and control of charge and spins in single dopants. However, an important problem remains, which is the control at the atomic scale of the dopants positioning. We propose to circumvent this problem by using 2 dimensional materials as hosts. Since the first isolation of graphene in 2004, the number of new 2D materials with favorable  properties for electronics has been growing. Dopants in 2 dimensional systems are more tightly bound and potentially easier to position and manipulate. Considering the properties of currently available 2D materials, we access the feasibility of such proposal in terms of the manipulability of isolated dopants (for single qubit operations) and dopant pairs (for two qubit operations). Our results indicate that a wide variety of 2D materials may perform at least as well as the currently studied bulk host for donor qubits.
\end{abstract}
\maketitle

Defects are an essential ingredient in semiconductor technology as they provide proper carriers to intrinsically insulating semiconductors. Dopants constitute the basis for transistor operations. The miniaturisation of these devices has moved defects to the forefront research, as their number and location may affect device performance and reproducibility~\cite{shinadaNat2005}. Few-donor specific configurations were explored  by Kane~\cite{KaneNature1998} in his Si quantum computer proposal, based on an array of donors in which each of them acts like a spin qubit. This in principle leads to a scalable quantum computer, and would be compatible with the existing Si-based transistor industry. For spin qubits, Si has the additional advantage of sustaining very long spin coherence times, up to seconds for isotopically purified Si~\cite{tyryshkinNatMater2012}.

\begin{figure}
\leavevmode
\includegraphics[clip,width=0.25\textwidth]{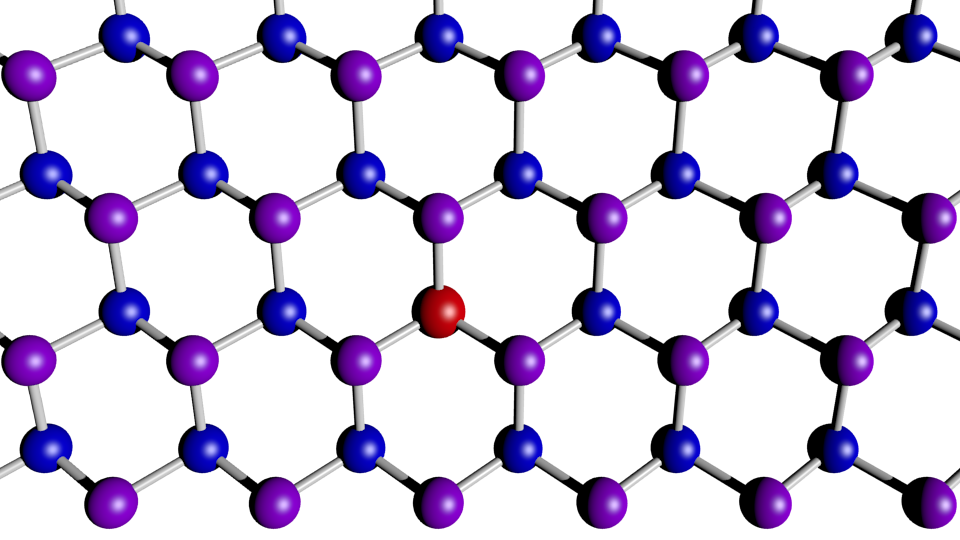}
\caption{Many of the 2D materials currently under study have a graphene like crystal structure (2 sublattices represented by blue and violet dots). Here we explore, within an effective mass approach, the possibility of using substitutional dopants (red dot) on such lattices to implement spin-qubits. Some of those 2D structures may present advantages over bulk (3D) semiconductor hosts.}
\label{fig:Fig1}
\end{figure}

The effort to understand single donor behavior has led to significant  raise of  expertise on the manipulation and control of states bound to donors in the last few years~\cite{morelloNat2010,plaNat2012,zwanenburgRMP2013,Gonzalez-ZalbaNanoLett2014,freerarXiv2016}. One problem of using donors in Si for qubits is that interference among the multiple degenerate Si conduction band minima states leads to a  sensitive and oscillatory behavior of tunnel~\cite{calderonJAP2009}  and exchange~\cite{koillerPRL2001}  coupling of electrons bound to pairs of donors as the relative positions of the donors vary.
Although no oscillatory behavior is expected for coplanar dopant pairs relative to (001) planes under tensile stress, any individual dopant deviation in the $z$ direction restores the oscillations~\cite{koillerPRB2002}.
This problem can be deterrent to quantum computing implementation in Si due to the relative lack of control on the exact position of dopants in the bulk.
Alternative proposals suggested to overcome this difficulty include hybrid dopant-quantum dot structures~\cite{Gonzalez-ZalbaNJP2012}, a charge-spin hybrid qubit~\cite{ShiPRL2012}, optical manipulation~\cite{abanto2009} and dipole coupling with electrons~\cite{TosiNatComm2017} or holes~\cite{SalfiPRL2016,Abadilloarxiv2017}. 

Here we propose an alternative which relies on 2 dimensional (2D) semiconductor materials instead of bulk Si for host material, as precise positioning of donors on a surface may be simpler than in the bulk, i.e., it involves control over two coordinates, avoiding the $z$-component uncertainties, see Fig.~\ref{fig:Fig1}. Moreover, many of the existing 2D materials present the conduction band minimum at $\Gamma$~\cite{OzcelikPRB2016,makPRL2010}, naturally getting rid of oscillatory exchange and tunnel couplings. 

The family of  2D  materials comprises an increasing number of elemental and compound semiconductors~\cite{CastellanosNatPho2016,novoselovSci2016,roldan2dimtheory}. Many have been experimentally isolated already, as research activity in this area raises. In the case of non-metallic behaviour their band gaps range from meV to a few eV. They can also be stacked in van der Waals heterostructures~\cite{geimNat2013,novoselovSci2016,FrisendaCSR2018} favoring miniaturization and device integration. Incorporation of dopants affects the properties of isolated or stacked monolayers~\cite{lin2dmat2016,FengNanoHoriz2017}, as they do in bulk systems. Here we explore doping in the very low density limit such that electrons can be bound to single and pairs of donors in a 2D environment in the context of quantum computation.

\begin{figure*}
\leavevmode
\includegraphics[clip,width=0.75\textwidth]{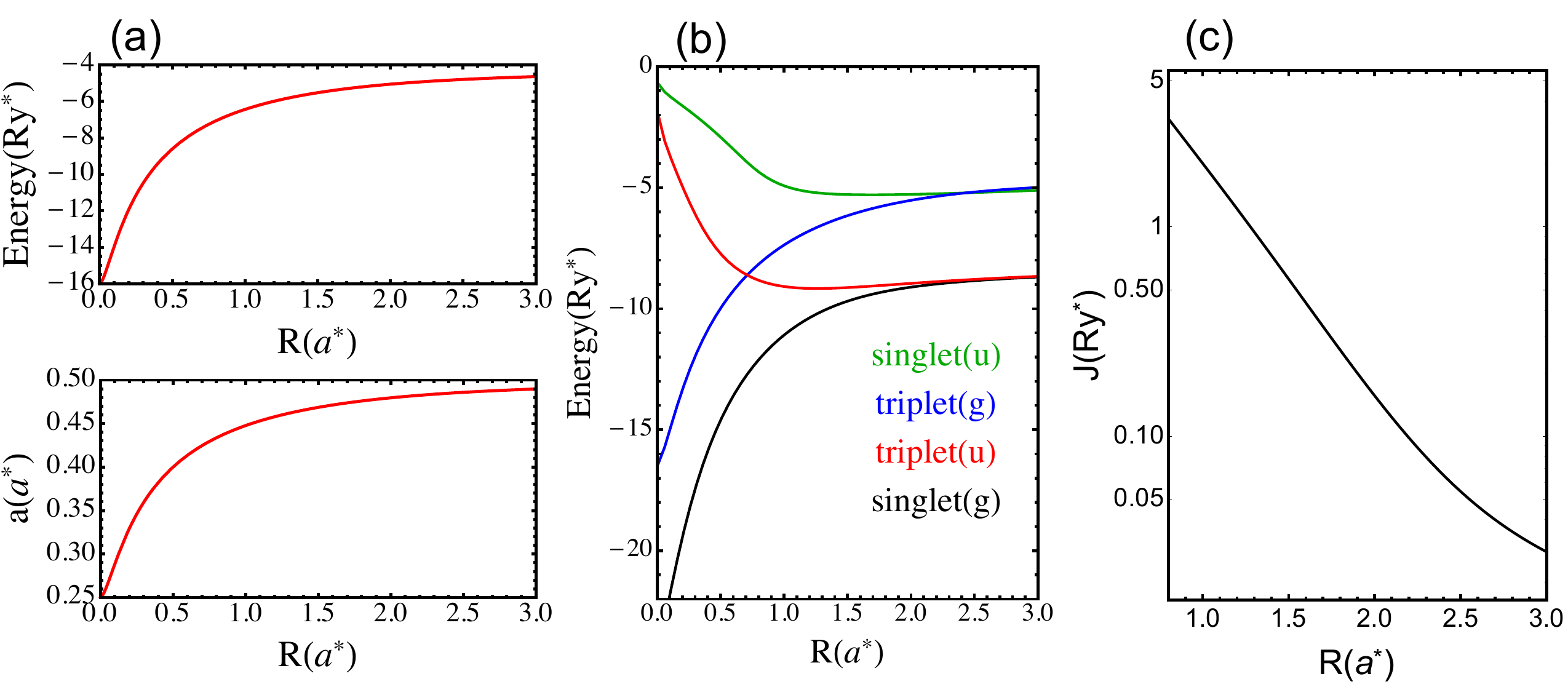}
\caption{(a) Bohr radii $a_{D_2^+}$ and energy for one electron bound to a donor pair $E_{D_2^+}$ as a function of $R$. For $R=2a^*$, $E_{D_2^+}=-5.06 Ry^*$ and $a_{D_2^+}=0.48 a^*$. Assuming $\epsilon=5$ and using the effective masses in Table I, $a_{D_2^+}^{\rm ZnS}=6.8$ \AA $\,$  and $E_{D_2^+}^{\rm ZnS}=-515$ meV, $a_{D_2^+}^{\rm CdS}=7.6$ \AA $\,$  and $E_{D_2^+}^{\rm CdS}=-460$ meV, $a_{D_2^+}^{\rm CdSe}=10$ \AA $\,$  and $E_{D_2^+}^{\rm CdSe}=-350$ meV, $a_{D_2^+}^{\rm SiC}=1.96$ \AA $\,$  and $E_{D_2^+}^{\rm SiC}=-1.77$ eV. Using $m_{\rm eff}$ and $\epsilon$ for MoS$_2$ and h-BN in Table I, we get $a_{D_2^+}^{\rm MoS_2}=2.2-2.7$ \AA $\,$  and $E_{D_2^+}^{\rm MoS_2}=-1.59$ eV,  and $a_{D_2^+}^{\rm h-BN}=0.5$ \AA $\,$  and $E_{D_2^+}^{\rm h-BN}=-15$ eV. (b) Energies for two electrons bound to a donor pair as a function of the inter-donor distance $R$, see Appendix for the wave-function definition.  (c) Exchange $J$ in effective units as a function of the separation between donors. For $R=2a^*$, $J=0.156 Ry^*$. For this distance, assuming $\epsilon=5$ and using the effective masses in Table I, $J_{\rm ZnS}=16$ meV, $J_{\rm CdS}=14$ meV, $J_{\rm CdSe}=11$ meV, $J_{\rm SiC}=55$ meV. Using $m_{\rm eff}$ and $\epsilon$ for MoS$_2$ and h-BN in Table I, we get $J_{\rm MoS_2}=50-60$ meV and $J_{\rm h-BN}=467$ meV. }
\label{fig:Fig2}
\end{figure*}

To this purpose, we analyze the stability of bound states in single dopants and the interaction between pairs of donors in a 2D semiconductor host
using an effective mass approach (EMA). We consider single donors and donor pairs in 2D. Within EMA the discrete crystal structure of the device is described by a continuum characterized by the effective mass $m_{\rm eff}$ and the dielectric screening $\epsilon$ of the host materials. In atomic units, the binding energy in 2D is larger than in 3D for a particular  $m_{\rm eff}$ and $\epsilon$. Defining the effective Rydberg as $Ry^*=\frac{m_{\rm eff}e^4}{2 \hbar^2 \epsilon^2}$, the binding energy of the electron bound to a single dopant in 3D is $E_B^{3D}=Ry^*$ while in 2D it is $E_B^{2D}=4Ry^*$. Similarly, defining $a^*=\frac{\hbar^2 \epsilon}{m_{\rm eff} e^2}$, the respective Bohr radii are $a^{3D}=a^*$ while $a^{2D}=a^*/2$. The values of the effective units depend on $m_{\rm eff}$ and $\epsilon$ as: $Ry^*=13.6 m_{\rm eff}/\epsilon^2$ eV and $a^*=0.529 \epsilon/m_{\rm eff}$ \AA. 

\begin{figure*}
\leavevmode
\includegraphics[clip,width=0.8\textwidth]{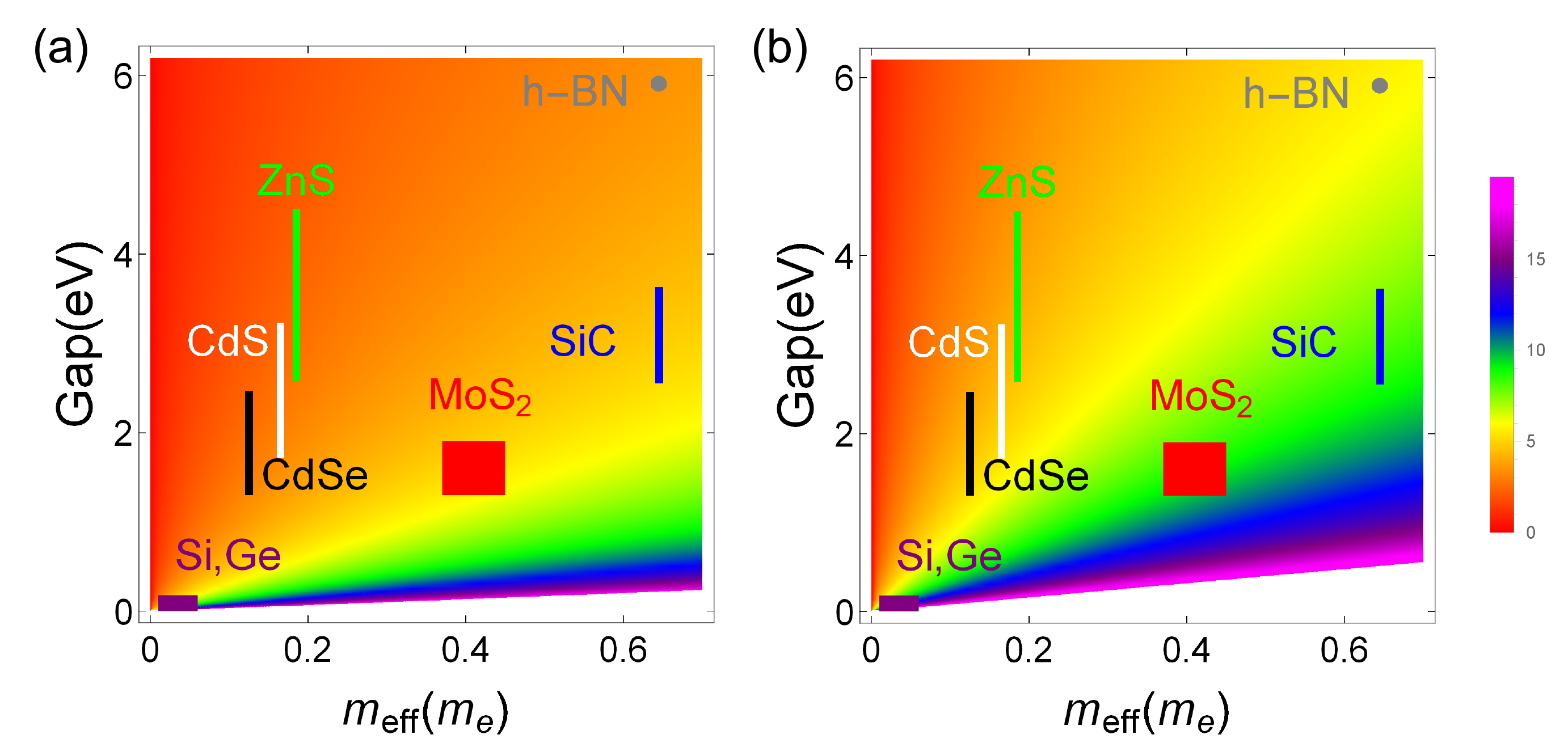}
\caption{Minimum dielectric constant that guarantees the existence of bound states and the validity of EMA for isolated dopants (a) and dopant pairs separated by $R=2a^*$ (b). Given the known values of $\epsilon$, we expect to be in the yellow-orange-red region of the map which encloses many of the analysed materials. The range of values for masses and gaps available in the literature and summarised in Table I are shown by the extended symbols next to the corresponding material composition. }
\label{fig:Fig3}
\end{figure*}
\begin{table*}
  \begin{tabular}{ | c | c | c | c | c | c |}
    \hline
    Material & Effective mass ($m_{\rm eff}$) & Gap (eV)  & Dielectric constant($\epsilon_0$)\\ \hline
   ZnS & 0.187 & 2.58-4.5 & - \\ \hline
   CdS & 0.167 & 1.72-3.23 & -\\ \hline
   CdSe & 0.127 & 1.30-2.47  & -\\ \hline
   SiC & 0.645 & 2.55-3.63 & - \\ \hline
   MoS$_2$ & 0.37~\cite{QiuPRL2013}-0.45~\cite{YoonNanoLett2011} & 1.3-1.9~\cite{CastellanosNatPho2016} & 4~\cite{SantosACSNano2013} \\ \hline
   h-BN & 1.175 & 5.9~\cite{CastellanosNatPho2016} &  2.31~\cite{LiNanoLett2015} \\ \hline
  \end{tabular}
   \caption{Effective masses and gaps of selected 2D materials. ZnS, CdS, CdSe and SiC have a direct gap at $\Gamma$. h-BN and MoS$_2$ have it at $K$. In the literature,  values for the dielectric constants (mostly calculated) can only be found for a few materials and, as discussed in the text, they depend on external conditions. Therefore, we consider the dielectric constant as a parameter. Unless otherwise stated, the data are taken from Ref.~\cite{MiroCSR2014}.}
\end{table*}

The gap and the effective masses of many different 2D materials have been estimated from band structure calculations~\cite{MiroCSR2014,OzcelikPRB2016}. Some gaps are also known experimentally from transport and optical measurements~\cite{CastellanosNatPho2016,novoselovSci2016,roldan2dimtheory}. The size and nature (direct or indirect) of the gap depends on the number of layers~\cite{WuNanoLett2015}, the distance between them~\cite{MiroCSR2014}, the nature of the substrate or atomic reconstructions~\cite{pflugradtPRB2014}. In some cases, it can be tuned with an electric field, as for the buckled silicene and germanene~\cite{NiNanoLett2012}. 

There is much less information on the dielectric screening of 2D materials, which also depends on the substrate and environment. It has been calculated only for a few cases (for instance, MoS$_2$~\cite{SantosACSNano2013} or h-BN~\cite{LiNanoLett2015}) and experimentally the reported values are very spread out~\cite{SantosACSNano2013}. Typically, the dielectric constant of monolayer materials is expected to be smaller than their 3D counterparts, as their screening capabilities are reduced in low dimensionality~\cite{LiNanoLett2015,WuNanoLett2015}. All this variability would give rise to an expected dispersion of the binding energy of dopants depending on external factors. Accordingly, it has been shown, using first principles calculations in transition metal dichalcogenides, that dopants can be tuned from deep to shallow by using different substrates~\cite{maPRB2017}. This modulation of ionization energy has been studied in the context of achieving p-type/n-type doping for transistor-like devices, but it certainly remains relevant for the donor quantum manipulation proposed here. 

Another important issue to take into account is the fact that in 2D systems the dielectric function is non-local. As discussed in Ref.~\cite{cudazzoPRB2011}, it may be written as $\epsilon({\bf q})=1+2\pi \alpha $, with $\alpha$ the polarizability. Hence for the description of the impurity potential we should take into account the dependence of the screening $\epsilon$ with  distance from the donor. However, it has been recently found that the effect of a non-local dielectric function can be reproduced by a dielectric constant  given by its average within the radius of the wave-function~\cite{olsenPRL2016}, dramatically simplifying the energy calculations. We consider this valid here, taking a constant $\epsilon$ to estimate binding energies.

We adopt isotropic envelopes, simplifying the calculations while keeping the physical picture~\cite{saraivaJPCM2015}. The hydrogenic 2D bound state is $\psi(r)=\sqrt{\frac{8}{\pi}} e^{-r/a}$ with $a=a^*/2$. For a single electron bound to a dopant pair $D_2^+$, see Appendix, the wave function radius and the binding energy are functions of the inter donor separation $R$, as shown in Fig.~\ref{fig:Fig2} (a). For $R=0$ one gets a He-like positive ion, He$^+$, with binding energy $16Ry^*$ and a Bohr radius $a^*/4$. For very large $R$, we recover the hydrogenic result, as the electron would only be bound to one of the dopants. For two electrons bound to a dopant pair $D_2$, at least two variational parameters are required, see Appendix.

The EMA is appropriate to describe shallow states in semiconductors, thus the gap of the considered material has to be much larger than the binding energies $E_B$. In order to implement this condition, we consider the generally unknown dielectric constant as a free parameter and estimate its minimum value required for the binding energy to fulfill the condition $E_B< E_g/2$ as a function of the gap $E_g$ and the effective mass on the conduction band, see Fig.~\ref{fig:Fig3}. Based on the known values for $\epsilon$, an estimate $\epsilon\lesssim5$ seems reasonable. This corresponds in the rainbow color code in Fig.~\ref{fig:Fig3} to the yellow-orange-red region of the plots. For shallow donors, the condition would be $E_B<< E_g/2$, and hence the points would be blue-shifted, meaning larger values of $\epsilon$. In order to put our results in the context of available 2D materials, we introduce in Fig.~\ref{fig:Fig3} data from Table I. The yellow-orange-red region of the plots includes various 2D materials which, in terms of energetics of the bound states, could host shallow donor states. The condition is somewhat more restrictive for donor pairs as the corresponding binding energy is enhanced for the short inter donor distances ($R=2a^*$) considered. For larger values of $R$, the small $\epsilon$ region is expanded. In general, in this yellow-orange-red region we find the first three materials in Table I, and possibly silicene and germanene if their gap is suitably enhanced.

In order to estimate binding energies and Bohr radii, we assume $\epsilon \sim 5$ for the first three materials in Table I. With this value, the Bohr radii for electrons bound to single dopants would be between $1$ and $2$ nm, which is comparable to the corresponding values for 3D Si~\cite{saraivaJPCM2015}. Consistently, the binding energies are similar to those in 3D silicon, with values ranging from $70$ to $100$ meV. The last three materials in Table I would be much more confined, with Bohr radii within a few \AA$\,$ and energies up to few eV.  Although EMA is not designed to treat large $E_B$ values, it is certain that the wave-functions in this limit are more confined (smaller effective Bohr radii). This is a desirable property in terms of isolation of the qubit and robustness against decoherence processes. 

Now we turn to the conditions for two qubit operations. In the original Si quantum computer proposal~\cite{KaneNature1998}, two-qubit operations are driven by exchange gates, i.e., exchange coupling $J$ pulses between electrons bound to neighboring donors. 

For 2 electrons bound to a single donor there are 2 low-energy levels well separated from the next excited state, one singlet  and one triplet, which allows to map the lower-energy states problem to the Heisenberg spin-1/2 Hamiltonian. For 2 electrons bound to a donor pair, there are 4 possible orbital states (see Appendix). 
We label  the  expectation values of the hamiltonian for these states in increasing order $E_1$, $E_2$, $E_3$ and $E_4$, and assign a spin hamiltonian to this problem if $E_2-E_1<<E_3-E_2$ such that only the two lowest levels are relevant at low temperatures, and the spin-1/2 hamiltonian may be defined as for a 2-level system. It can be shown that the 2 lowest levels are a singlet and a triplet, as illustrated in Fig.~\ref{fig:Fig2}(a). 
Fig.~\ref{fig:Fig2}(c) shows $J$,  the difference between the lowest singlet and triplet levels, vs $R$ in a physically accessible range of inter donor distances. For $R=2a^*$, $J=0.156 Ry^*$.  For $\epsilon\sim 5$ and interdonor separation $R=2a^*$, the exchange values cover a wide range, $15$ meV $< J(R=2a^*) < 100$ meV, for the materials in the yellow-orange-red region in Fig.~\ref{fig:Fig3}. With these $J$ values one would perform very fast (about $10^{-14}$ s) manipulations for $\sqrt{\rm SWAP}$ operations. If coherence times in 2D are about the same as in 3D, this would allow a large number of operations within coherence times.  $J$ can be strongly enhanced in materials with larger binding energies but, in this case, sub-nm inter dopant distances would be required, demanding a very high accuracy for the placement of gates on top and between donors.

In conclusion, the variability of binding energies as a function of the chemical composition, substrate and number of layers, opens up a wide range of possibilities for the potential use of dopants in 2D materials for quantum computation. We distinguish 2D materials that support shallow states with binding energies and Bohr radii comparable to $P$ in Si, and those that support stronger confinement. Each group of materials could serve different
purposes with shallower states more suitable for manipulation and deeper ones for storage. The synergy among different experimental techniques for dopant positioning in 3D semiconductors, combined with recent advances in 
2D materials-based electronics and multilayered architectures control, provide key technical tools for the practical implementation of donor-based spin qubits.

{\it Acknowledgements}. JCAU and MJC acknowledge funding from Ministerio de Econom\'ia, Industria y Competitividad (Spain) via Grants No FIS2012-33521 and FIS2015-64654-P and from CSIC (Spain) via grant No 201660I031 JCAU thanks the support from grant BES-2013-065888. In Brazil (BK) this work is part of the Brazilian National Institute for Science and Technology on Quantum Information. BK also acknowledge partial support from FAPERJ, CNPq.

{\bf Appendix: Variational wave-functions}.

We are interested here in the ground state properties of 1 or 2 electrons bound to single donors or donor pairs, therefore a variational scheme is appropriate. For non-degenerate conduction band edge the rescaled atomic H ground state is a good trial wavefunction for one-donor situations. In case of 2 donors, a properly symmetrized combination of hydrogenic orbitals centered at each donor is a reasonable choice for trial wavefunction.   
For example, for the molecular  ion D$_2^+$ ground state we take the trial form $\Psi=N \left(e^{ -r_A/a'}+e^{-r_B/a'} \right)$ where $r_A$ and $r_B$ represent the electron's distance from the donors  $A$ and $B$ and $a'$ is a variational parameter, chosen to minimize the expectation value of the energy.                                                                                                                               
For the neutral molecule D$_2$ we take the combinations 

\begin{widetext}
\begin{eqnarray}
\Psi^g_{singlet}(\mathbf{r}_1,\mathbf{r}_2)=N [ (e^{-\alpha r_{1A}-\beta r_{2B}}+ e^{-\alpha r_{2A}-\beta r_{1B}})+(e^{-\alpha r_{1B}-\beta r_{2A}} + e^{-\alpha r_{2B} -\beta r_{1A}}) \nonumber \\ +e^{i\phi}(e^{-\alpha (r_{1A}+r_{2B})}+e^{-\alpha (r_{2A}+r_{1B})})+e^{i\theta}(e^{-\beta (r_{1A}+r_{2B})}+e^{-\beta (r_{2A}+r_{1B})})]
\label{Eq:2parsin}
 \\
\Psi^{u}_{singlet}(\mathbf{r}_1,\mathbf{r}_2)=N [ (e^{-\alpha r_{1A}-\beta r_{2B}}+ e^{-\alpha r_{2A}-\beta r_{1B}})-(e^{-\alpha r_{1B}-\beta r_{2A}} + e^{-\alpha r_{2B} -\beta r_{1A}})]
\label{Eq:2partsu}
 \\
\Psi^{g}_{triplet}(\mathbf{r}_1,\mathbf{r}_2)=N [ (e^{-\alpha r_{1A}-\beta r_{2B}}- e^{-\alpha r_{2A}-\beta r_{1B}})+(e^{-\alpha r_{1B}-\beta r_{2A}} - e^{-\alpha r_{2B} -\beta r_{1A}})]
\label{Eq:2partg} \\
\Psi^{u}_{triplet}(\mathbf{r}_1,\mathbf{r}_2)=N [ (e^{-\alpha r_{1A}-\beta r_{2B}}- e^{-\alpha r_{2A}-\beta r_{1B}})-(e^{-\alpha r_{1B}-\beta r_{2A}} - e^{-\alpha r_{2B} -\beta r_{1A}}) \nonumber \\ +e^{i\phi}(e^{-\alpha (r_{1A}+r_{2B})}-e^{-\alpha (r_{2A}+r_{1B})})+e^{i\theta}(e^{-\beta (r_{1A}+r_{2B})}-e^{-\beta (r_{2A}+r_{1B})})]
\label{Eq:2partu}
\end{eqnarray}
\end{widetext}
$\alpha$ and $\beta$ are the variational parameters. These variational wave-functions correspond to the four lowest bound states. Only the orbital part is explicitly given -  the spin part is inferred by symmetry and is indicated by the label singlet or triplet. The combination of different terms is included to preserve the symmetry (symmetric: gerade or antisymmetric: ungerade) under reflection $A\leftrightarrow B$. The undetermined relative phases $\theta$ and $\phi$, included to allow for the most generic wave-functions that fulfill the symmetry of H$_2$, are found to be both zero in the minimization process.

\bibliography{bibtex2dim}
\end{document}